\documentclass[a4paper]{revtex4}

\usepackage{amssymb,amsmath,amsfonts}
\usepackage{mathrsfs}
\usepackage{dsfont}
\usepackage{graphicx,bm}
\usepackage{color}

\begin{document}

\title{Hybrid star properties from an extended linear sigma model}

\author{J\'anos Tak\'atsy, P\'eter Kov\'acs and Gy\"orgy Wolf}

\affiliation{Institute for Particle and Nuclear Physics, Wigner
  Research Center for Physics, Hungarian Academy of Sciences, H-1525
  Budapest, Hungary}

\email{wolf.gyorgy@wigner.hu}

\begin{abstract} 
The equation of state provided by effective models of strongly interacting matter should comply with the restrictions imposed by current astrophysical observations on compact stars. Using the equation of state given by the (axial-)vector meson extended linear sigma model, we determine the mass-radius relation and study whether these restrictions are satisfied under the assumption that most of the star is filled with quark matter. We study the dependence of the mass-radius relation on the parameters of the model.
\end{abstract}


\maketitle

\section{Introduction}

By the recent astrophysical observation of heavy neutron stars \cite{Demorest:2010} and neutron star collisons \cite{GW170817:2018} the interest for studying dense strongly interacting matter increased substantially. In terrestial experiments ALICE at CERN, and PHENIX and STAR at RHIC \citep{Tlusty:2018rif} explored the strongly interacting matter at low density and high temperature. In this case, the situation is also satisfactory on the theoretical side since lattice calculations are applicable. At high density the theoretical investigations are not at such a satisfactory level, therefore, effective models are needed in this region where the existing experimental data (NA61 at CERN, BES/STAR at RHIC) are scarce and have rather bad statistics. Soon to be finished experimental facilities (NICA at JINR and CBM at FAIR) are designed to explore this region more precisely \cite{Kekelidze:2014dta,Ablyazimov:2017guv}.

Neutron stars are on the one side a challenge requiring to understand their properties i.e mass, radius, tidal deformability, structure etc. On the other hand, they can provide information about the dense, strongly interacting matter in a region of the phase diagram that is inaccessible to terrestrial experiments \cite{lattimer2007,Watts:2016uzu,Tews:2018chv}. Since the Tolman-Oppenheimer-Volkoff (TOV) equations \cite{Tolman, OppVolk} provide a direct relation between the equation of state (EoS) of the compact star matter and the mass-radius (M-R) relation of the compact star, these data can help to select those effective models, used to describe the strongly interacting matter, whose predictions are consistent with compact star observables. For example, the EoS must support the existence of a two-solar-mass neutron star with star radii in the permitted radius window of 11-12.5km, {\it cf.} \cite{Hell:2014xva,lattimer2014}.

Based on the above considerations, we investigate mass-radius sequences given by the EoS obtained for the quark phase in \cite{kovacs2016} from the $N_f=2+1$ flavor extended linear sigma model introduced in \cite{parganlija2013}. For the hadronic phase we use for the outer crust the BPS EoS \cite{Baym:1971pw}, for the inner crust the NV EoS \cite{Negele:1973} and for the outer core until the phase transition we use the TNI2u \cite{TNIu:2006} EoS.  The model used here should be regarded as a rather crude approximation since the hadron-quark phase transition is introduced  in a very ad hoc way. The present work has to be considered only as our first attempt to study the problem.

The paper is organized as follows. In Section~\ref{sec:model} we present the model and discuss how its solution was obtained in \cite{kovacs2016}, which reproduced quite well some thermodynamic quantities measured on the lattice. It can be used in the presence of a vector meson introduced here to realize the short-range repulsive interaction between quarks in the simplest possible way. In Section~\ref{sec:res} we compare our results for the EoS and the M-R relation (star sequences), for various values of the unknown model parameters, such as the vector coupling $g_v$, the quark-hadron crossover transition density $\bar\rho$. We draw the conclusions in Section~\ref{sec:concl} and discuss possible ways to improve the treatment of the model.

\section{Methods} \label{sec:model}

For the EoS of the hadronic phase we use models from nuclear physics. For very low density, (outer crust, $\rho< 0.01 \rho_0$) we use the BPS EoS \cite{Baym:1971pw} that corresponds to a Coulomb lattice of different nuclei embedded in a gas of electrons and that has a remarkable influence on the mass-radius relation of the star.

For the region of $0.01\rho_0 \le \rho < \rho_0$ (inner crust) we use the NV EoS \cite{Negele:1973} and above this until the phase transition we use the TNI2u or TNI3u \cite{TNIu:2006,Masuda:2013} EoS, which differ in their stiffness (K=250 MeV or K=300 MeV, respectively). These EoS's even with their derivatives go over each other continuously, so there is no need of any treatment at the connections and the exact density of the transitions are irrelevant. The model used in this paper for the quark phase is an $N_f = 2+1$ flavor (axial)vector meson extended linear sigma model (eLSM). The Lagrangian and the detailed description of this model, in which in addition to the full nonets of (pseudo)scalar mesons the nonets of (axial)vector mesons are also included, can be found in \cite{kovacs2016,parganlija2013}. The model contains three flavors of constituent quarks, with kinetic terms and Yukawa-type interactions with the (pseudo)scalar mesons. An explicit symmetry breaking of the mesonic potentials is realized by external fields, which results in two scalar expectation values, $\phi_\mathrm{N}$ and $\phi_\mathrm{S}.$ 

Compared to \cite{kovacs2016}, the only modification to the model is that we include in the Lagrangian a Yukawa term $-g_v\sqrt{6}\bar\Psi \gamma_\mu V_0^\mu\Psi$, which couples the quark field $\Psi^{\rm T}=(u,d,s)$ to the $U_V(1)$ symmetric vector field, that is $V_0^\mu = \frac{1}{\sqrt{6}} \mathrm{diag}(v_0 + \frac{v_8}{\sqrt{2}}, v_0 + \frac{v_8}{\sqrt{2}}, v_0-\sqrt{2}v_8)^\mu$. The vector meson field is treated at the mean-field level as in the Walecka model \cite{walecka}, but as a simplification we assign a nonzero expectation value only to $v_0^0$: $v_0^\mu\to v_0\delta^{0\mu}$ and $v_8^\mu\to 0$. In this way the chemical potentials of all three quarks are shifted by the same amount, allowing us to use, as shown below, the result obtained in \cite{kovacs2016}. With the parameters used in \cite{kovacs2016}, the mass of the vector meson $v_0^\mu$ turns out to be $m_v=871.9$~MeV.

Since a compact star is relatively cold ($T\approx 0.1$~keV), we work at $T=0$~MeV. We have three background fields $\phi_\mathrm{N},\phi_\mathrm{S}$ and $v_0$, and the calculation of the grand potential, $\Omega$, is performed using a mean-field approximation, in which fermionic fluctuations are included at one-loop order, while the mesons are treated at tree-level. Hence, the grand potential can be written in the following form:
\begin{eqnarray}
\label{Eq:grand_pot}
\Omega(\mu_q; &&\!\!\!\phi_\mathrm{N}, \phi_\mathrm{S},v_0) = U_\mathrm{mes}(\phi_\mathrm{N},\phi_\mathrm{S}) - \frac{1}{2}m_v^2 v_0^2 \nonumber\\
&& + \Omega^{(0)\mathrm{vac}}_{q\bar{q}}(\phi_\mathrm{N}, \phi_\mathrm{S}) 
    + \Omega^{(0)\mathrm{matter}}_{q\bar{q}}(\tilde\mu_q;\phi_\mathrm{N}, \phi_\mathrm{S}) \:,
\end{eqnarray}
where $\tilde\mu_q=\mu_q-g_v v_0$ is the effective chemical potential of the quarks, while $\mu_q = \mu_\mathrm{B}/3$ is the physical quark chemical potential, with $\mu_B$ being the baryochemical potential. On the right hand side of the grand potential \eqref{Eq:grand_pot}, the terms are (from left to right): the tree-level potential of the scalar mesons, the tree-level contribution of the vector meson, the vacuum and the matter part of the fermionic contribution at vanishing mesonic fluctuating fields. The fermionic part is obtained by integrating out the quark fields in the partition function. The vacuum part was renormalized at the scale $M_0=351$~MeV. More details on the derivation can be found in \cite{kovacs2016}.

The background fields $\phi_\mathrm{N}, \phi_\mathrm{S}$, and $v_0$ are determined from the stationary
conditions
\begin{equation}
    \frac{\partial \Omega}{\partial \phi_\mathrm{N}}\bigg|_{\phi_\mathrm{N}=\bar\phi_\mathrm{N}} = \frac{\partial \Omega}{\partial \phi_\mathrm{S}}\bigg|_{\phi_\mathrm{S}=\bar\phi_\mathrm{S}}  = 0 \quad\textnormal{and} \quad \frac{\partial \Omega}{\partial v_0}\bigg|_{v_0=\bar{v}_0} =0 \: ,
    \label{Eq:stat_cond}
\end{equation}
where the solution is indicated with a bar. Since $\partial/\partial v_0=-g_v \partial/\partial{\tilde \mu_q}$, the stationary condition with respect to $v_0$ reads 
\begin{equation}
\label{Eq:gap-w0}
\bar{v}_0(\phi_\mathrm{N},\phi_\mathrm{S}) = \frac{g_v}{m_v^2} \rho_q(\tilde\mu_q(\bar{v}_0);\phi_\mathrm{N},\phi_\mathrm{S}) \:,
\end{equation}
where $\rho_q(x;\phi_\mathrm{N},\phi_\mathrm{S})=-\partial\Omega^{(0)\mathrm{matter}}_{q\bar{q}}(x;\phi_\mathrm{N},\phi_\mathrm{S})/\partial x.$

 When solving the model, the value of $g_v$ does not effect the properties of the system at zero chemical potential, therefore, we cannot fit it to the vacuum properties, so we leave it in the range of $[0,3)$ as a free parameter, while for the remaining 14 parameters of the model Lagrangian we use the values given in Table IV of \cite{kovacs2016}. These values were determined there by calculating constituent quark masses, (pseudo)scalar curvature masses and decay widths at $T=\mu_q=0$ and comparing them to their experimental PDG values \cite{PDG}. Parameter fitting was done using a multiparametric $\chi^2$ minimization procedure \cite{james1975}. In addition to the vacuum quantities, the pseudocritical temperature $T_\mathrm{pc}$ at $\mu_q=0$ was also fitted to the corresponding lattice result \cite{aoki2006, Bazavov:2011nk}. We mention that the model also contains the Polyakov-loop degrees of freedom (see \cite{kovacs2016} for details), but to keep the presentation simple we omitted them from \eqref{Eq:grand_pot}, as at $T=0$ they do not contribute to the EoS directly. Their influence is only through the value of the model parameters taken from \cite{kovacs2016}: since they influence the value of $T_\mathrm{pc}$ used for parameterization.
 
\begin{figure*}
\centerline{\includegraphics[width=0.49\linewidth]{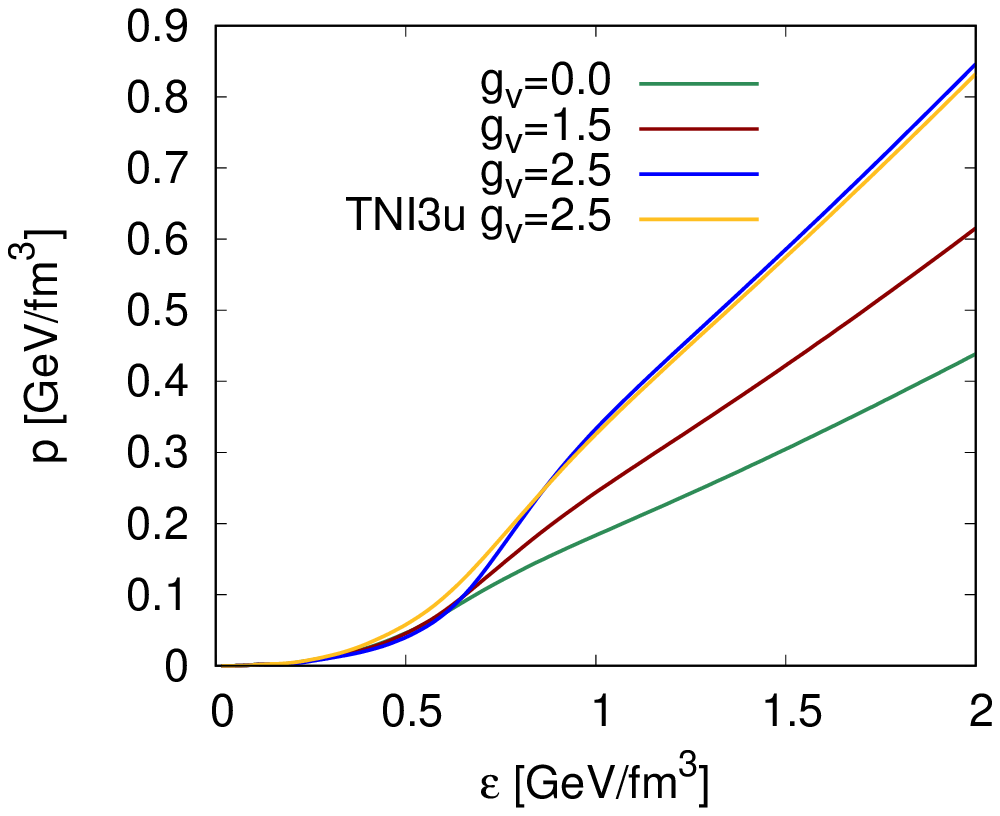}
\includegraphics[width=0.49\linewidth]{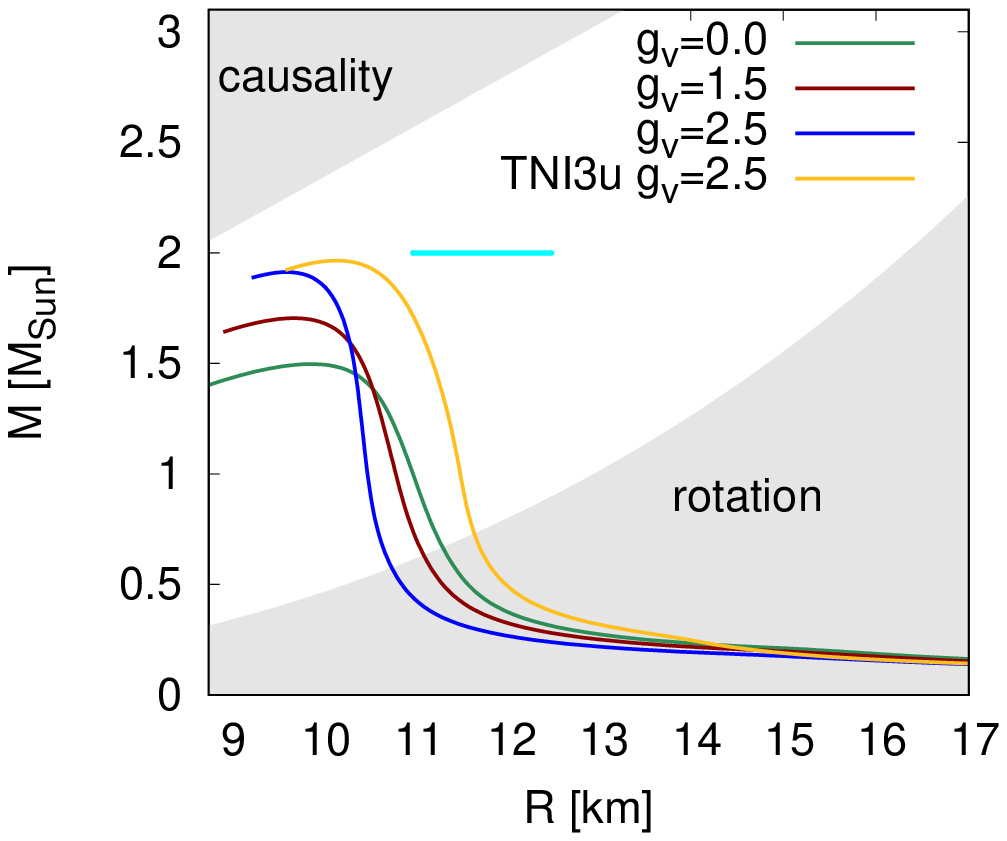}}
\caption{The equation of state (left) and the M-R curves (right) for different values of $g_v$ using $\bar\rho =4 \rho_0$ and $\Gamma=1$ with TNI2u (K=250 MeV) nuclear force for different $g_v$'s. The $g_v=2.5$ case is calculated with TNI3u (K=300 MeV), as well. In the right figure the observational constraint set by observed pulsars for the permitted radius window of $11.0$-$12.5$~km at $2$~$M_\odot$ ({\it e.g.} \cite{Hell:2014xva,lattimer2014}) is represented by the horizontal line. The different shaded regions are excluded by causality $R>2.9GM/c^2$ and the rotational constraint based on the $716$~Hz pulsar J1748-2446ad, $M/M_\odot>4.6\cdot10^{-4} \: (R/\mathrm{km})^3$ \cite{hessels2006}}
\label{fig:MR_gv}
\end{figure*}

 The solution of the model at $g_v=0$, obtained in \cite{kovacs2016}, can be used to construct the solution at $g_v\ne0$ (see {\it e.g.} Ch. 2.1 of \cite{almasi}): one only has to interpret the solution at $g_v=0$ as a solution obtained at a given $\tilde \mu_q$ and determine $\mu_q$ at some $g_v\ne 0$ using \eqref{Eq:gap-w0}. To see that the solutions $\bar\phi_\mathrm{N,S}$ for $g_v\ne0$ can be related to the solution obtained at $g_v=0,$ where $\bar{v}_0=0$, consider the grand potential at $g_v=0$. This potential, denoted as $\Omega_0$, is subject to the stationary conditions \eqref{Eq:stat_cond} with solutions $\bar\phi_\mathrm{N,S}^0(\mu_q)$. It is then easy to see using \eqref{Eq:grand_pot}, that the solution $\bar\phi_\mathrm{N,S}(\mu_q)$ of \eqref{Eq:stat_cond} satisfies $\bar\phi_\mathrm{N,S}(\mu_q+g_v v_0)=\bar\phi_\mathrm{N,S}^0(\mu_q)$ or, changing the variable $\mu_q$ to $\tilde \mu_q$ the relation becomes 
\begin{equation}
\bar\phi_\mathrm{N,S}(\tilde\mu_q+g_v v_0)=\bar\phi_\mathrm{N,S}^0(\tilde\mu_q).
\end{equation}
The value of the grand potential $\Omega$ at the extremum can be given in terms of the value of the grand potential with $g_v=0$, that is $\Omega_0$, at its extremum. Using that the extrema of $\Omega_0(\tilde\mu_q,\phi_\mathrm{N},\phi_\mathrm{S},v_0=0)$ are $\bar\phi_\mathrm{N}^0$ and $\bar\phi_\mathrm{S}^0$, one has
 \begin{eqnarray}
  \Omega(\mu_q;&&\bar\phi_\mathrm{N}(\mu_q), \bar\phi_\mathrm{S}(\mu_q), \bar{v}_0)
  \nonumber \\
  &&= \Omega_0(\tilde\mu_q,\phi_\mathrm{N}^0(\tilde\mu_q),\phi_\mathrm{S}^0(\tilde\mu_q),v_0=0)-\frac{1}{2}m^2_v\bar{v}_0^2\: ,
\end{eqnarray}
where
\begin{eqnarray}
  \bar{v}_0&\equiv&\bar{v}_0(\bar\phi_\mathrm{N}(\mu_q),\bar\phi_\mathrm{S}(\mu_q))=\frac{g_v}{m^2_v} \rho_q(\tilde\mu_q; \bar\phi_\mathrm{N}(\mu_q), \bar\phi_\mathrm{S}(\mu_q)) \nonumber \\
  &=&\frac{g_v}{m^2_v} \rho_q(\tilde\mu_q; \bar\phi_\mathrm{N}^0(\tilde\mu_q), \bar\phi_\mathrm{S}^0(\tilde\mu_q))
\end{eqnarray}
and $\mu_q=\tilde\mu_q+g_v \bar{v}_0$.
The pressure $p$ and the energy density $\varepsilon$ are calculated from the grand potential. At $v_0\ne0$ they can be expressed in terms of the pressure obtained at $g_v=0$: 
\begin{eqnarray}
    p(\mu_q) &=& \Omega(\mu_q=0;\bar\phi_\mathrm{N}(0),\bar\phi_\mathrm{S}(0),\bar{v}_0(0)) 
    - \Omega(\mu_q; \bar\phi_\mathrm{N}, \bar\phi_\mathrm{S},\bar{v}_0)\nonumber\\
    &=& \Omega_0(\tilde\mu_q=0;\bar\phi_\mathrm{N}^0(0),\bar\phi_\mathrm{S}^0(0),v_0=0) \nonumber \\
    &&- \Omega_0(\tilde\mu_q; \bar\phi_\mathrm{N}^0, \bar\phi_\mathrm{S}^0,v_0=0)
    +\frac{1}{2} m^2_v\bar{v}_0^2\nonumber\\ 
    &=& p(\tilde\mu_q)|_{g_v=0} + \frac{1}{2} m^2_v\bar{v}_0^2\: ,
\end{eqnarray}
where $\bar{v}_0=\frac{g_v}{m^2_v} \rho_q(\tilde\mu_q; \bar\phi_\mathrm{N}^0(\tilde\mu_q), \bar\phi_\mathrm{S}^0(\tilde\mu_q))$, and then $\varepsilon = -p + \mu_q \rho_q$, where $\mu_q=\tilde\mu_q+g_v\bar{v}_0$.

\begin{figure}
\centerline{\includegraphics[width=0.49\linewidth]{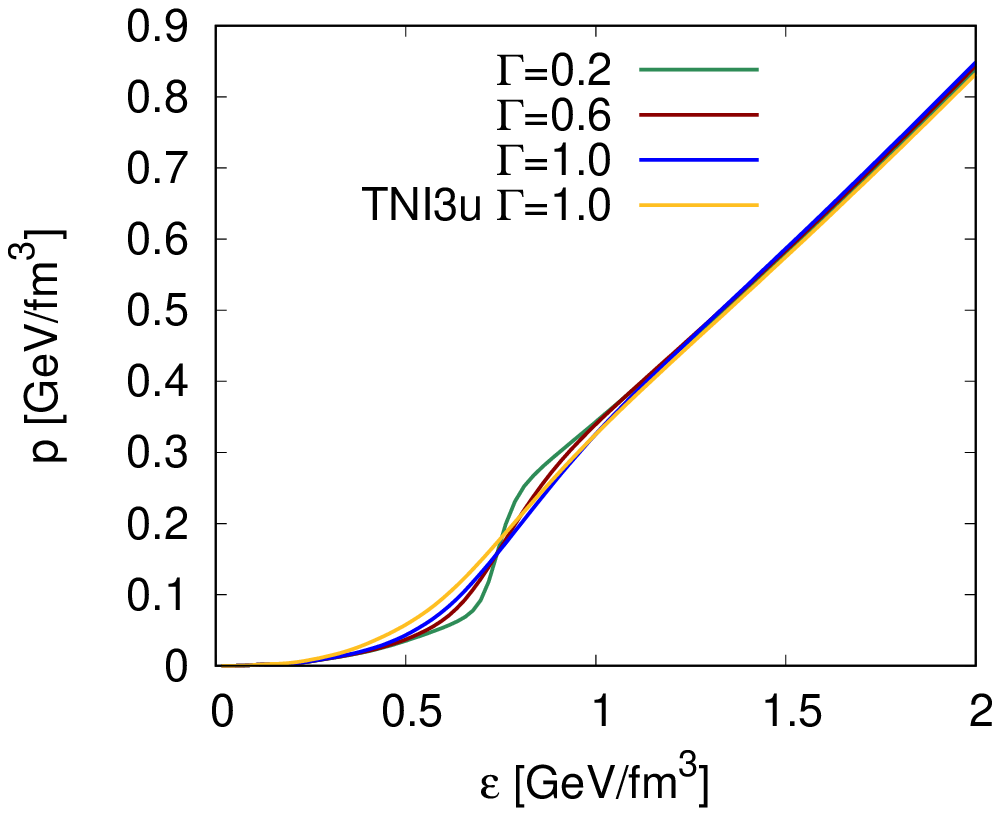}
\includegraphics[width=0.49\linewidth]{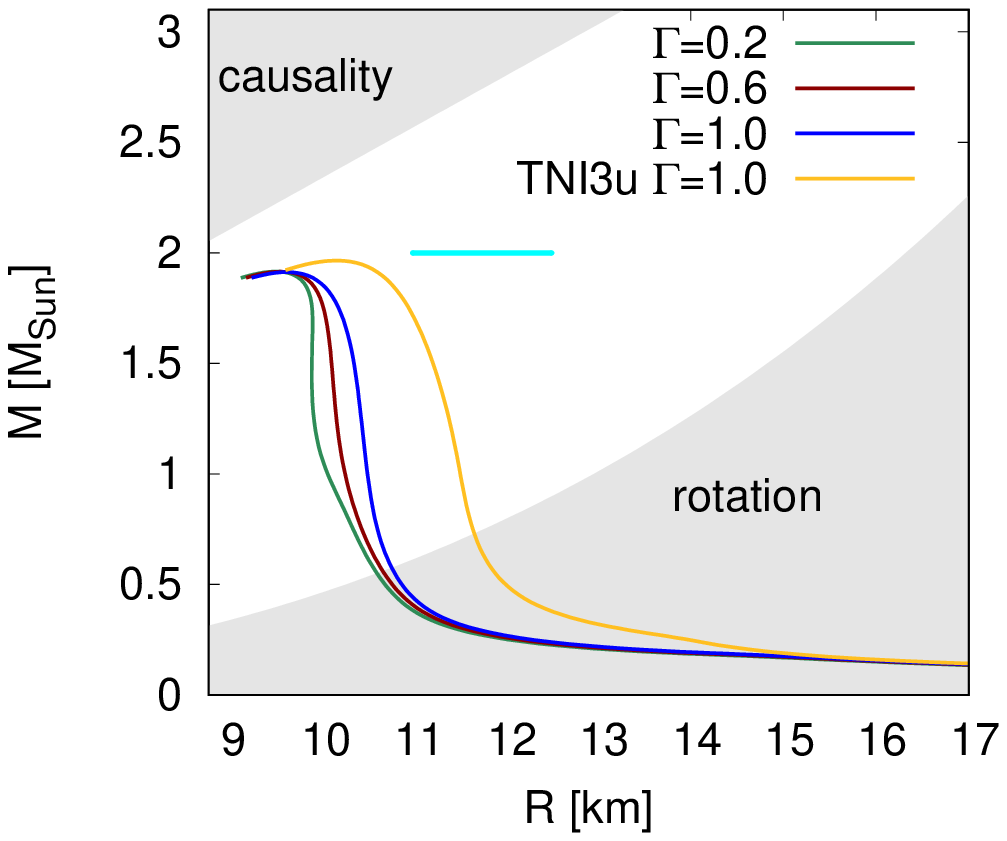}}
\caption{The equation of state (left) and the M-R curves (right) using $\bar\rho =4 \rho_0$ and $g_v=2.5$ with TNI2u (K=250 MeV) nuclear force for different $\Gamma$'s. The $\Gamma=1.0$ case is calculated with TNI3u (K=300 MeV), as well. The shaded regions in the right figure are explained in Fig.\ref{fig:MR_gv}.}
\label{fig:MR_Gamma}
\end{figure}

To connect the nuclear phase (TNI2u) and the quark phase EoS we assume a
hadron-quark crossover, and use the $P$-interpolation \cite{Asakawa:1997}:
\begin{eqnarray}
    P(\rho) &=&P_H(\rho) f_-(\rho)+P_Q(\rho) f_+ (\rho), \\
    f_{\pm}(\rho) &=& \frac{1}{2} \left( 1\pm \mathrm{tanh}\left( \frac{\rho-\bar{\rho}}{\Gamma} \right) \right)
  \end{eqnarray}
  \begin{eqnarray}
    \varepsilon (\rho)
    &=&\varepsilon_H (\rho) f_- (\rho)+\varepsilon_Q (\rho) f_+(\rho) 
    +\Delta \varepsilon  \\
    \Delta \varepsilon&=&\rho \int^{\rho}_{\bar{\rho}}(\varepsilon_H(\rho')
    -\varepsilon_Q(\rho'))\frac{g(\rho')}{\rho'}d\rho' \nonumber \\
    g(\rho')&=& \frac{1}{2\Gamma} \mathrm{cosh}^{-2}\left( \frac{\rho'-\bar\rho}{\Gamma} \right) .
  \end{eqnarray}
where $\bar\rho$ is the transition density and $\Gamma$ is the width of the
crossover transition. Note that neither the transition density - $\bar\rho$ -
nor the order of the phase transition between the hadron and the quark phases
are known, therefore, we will investigate the dependence of the M-R curves
on these parameters, as well. 

By solving the TOV equation  \cite{Tolman,OppVolk} using a fourth-order Runge-Kutta differential equation integrator with adaptive stepsize-control for a specific EoS, one can obtain the radial dependence of the energy density (and thus of the pressure) for a certain central energy density, $\varepsilon_0$. One can then determine the mass and radius of the compact star for that central energy density. By changing $\varepsilon_0$, one gets a sequence of compact star masses and radii parameterized by the central energy density, as shown in Fig.~\ref{fig:MR_gv} for various models. The sequence of stable compact stars ends when the maximum compact star mass is reached with increasing central energy density. 


\section{Results}  \label{sec:res}

As described above we use an equation of state (EoS) sewn together from 3 nuclear (BPS, NV and TNIu) and from the EoS for the quark phase. We consider that the low density (below $\rho< \rho_0$) nuclear EoS's are safe, their ingredients are based on well known and tested nuclear physics data, and their slight modifications change only the propeties of the crust, and do not show up significantly in those neutron star properties which we study here. Here we investigate the effect of the not very well determinable parameters $\bar\rho, \Gamma, g_v$ on the M-r curves and on the tidal deformability. We study moreover, what the effect of the stiffnes of the hadronic EoS is, so we compare results using TNI3u (K=300 MeV) and TNI2u (K=250 MeV) interactions.

\begin{figure}
\centerline{\includegraphics[width=0.49\linewidth]{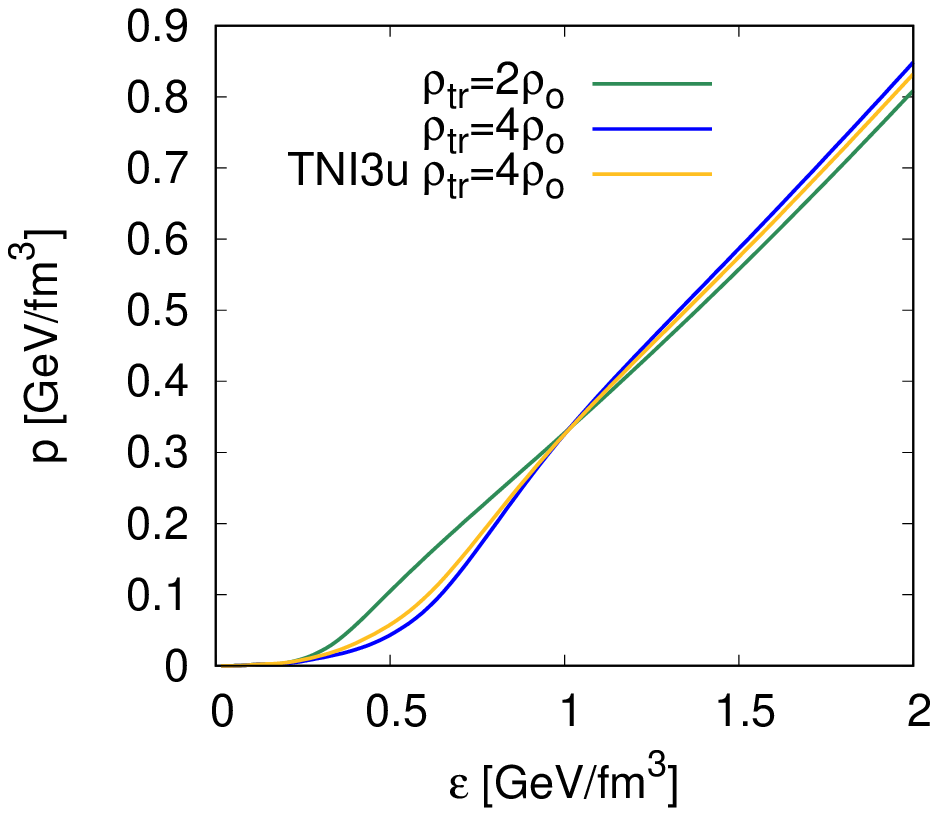}
\includegraphics[width=0.49\linewidth]{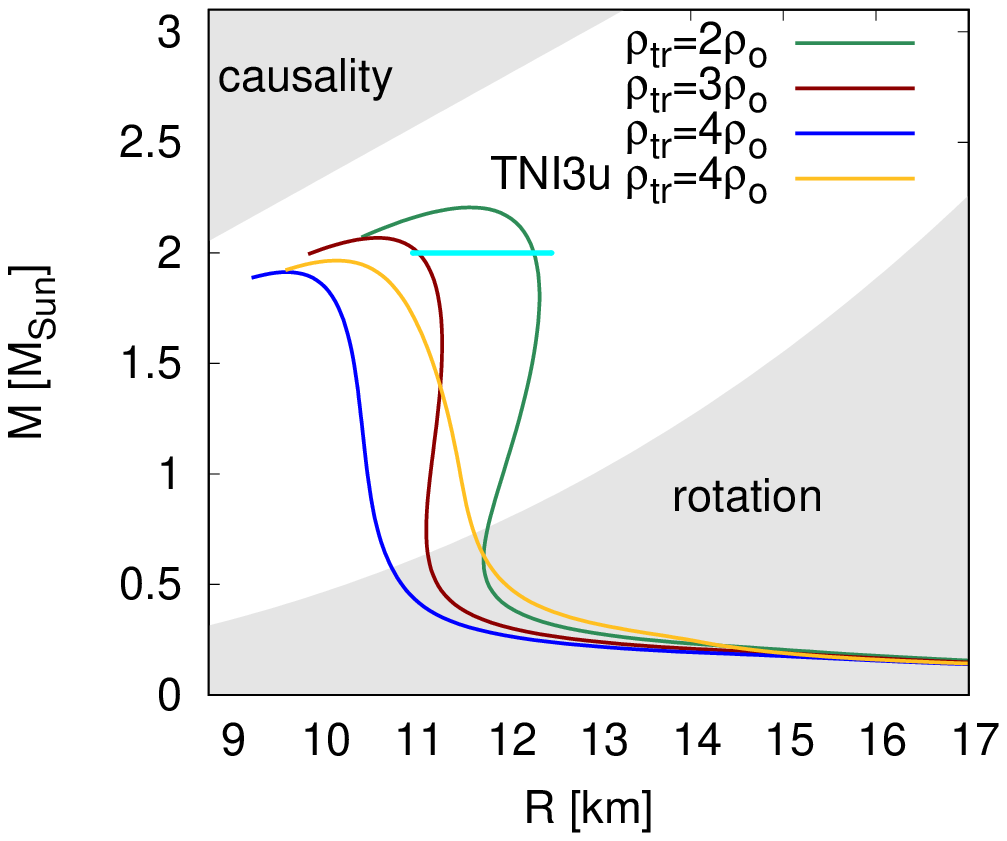}}
\caption{The equation of state (left) and the M-R curves (right) for different values of $\bar\rho$ using $g_v =2.5$ and $\Gamma=1$ with TNI2u (K=250 MeV) nuclear force for different $\bar\rho$'s. The $\bar\rho=4\rho_0$ case is calculated with TNI3u (K=300 MeV), as well. The shaded regions are explained in Fig.\ref{fig:MR_gv}.}
\label{fig:MR_rho}
\end{figure}

Inclusion of the repulsive interaction between quarks in the eLSM renders the EoS stiffer compared to the $g_v=0$ case, as expected, and at lower densities one can observe the effect of the stiffnes of the hadronic phase. It is worth noting that relatively small differences in the $p(\epsilon)$ lead to significant differences in the M-R curves, as we shall see in Figs.~\ref{fig:MR_gv}, \ref{fig:MR_Gamma} and \ref{fig:MR_rho}. It can also be seen that the maximum possible compact star mass is lower for models with less stiff EoSs. The highest mass compact star has a mass of $\sim1.5$~$M_\odot$ with $g_v=0$. The effect of hadronic compressibility is rather large, and the higher stiffnes brings the M-R curve closer to the observed radius window.

The effect of the crossover width $\Gamma$ is sizeable for either on the EoS or in the M-R sequence \ref{fig:MR_Gamma}, however, it does not effect the maximal neutron star mass and the corresponding radius. The EOS for $\Gamma=0.2$ is acausal ($c_s> 1$) around the transition point, so that solution is not physical.

As it could be expected the variation of transition density $\bar\rho$ has a large effect on the EoS and on the M-R curve (Fig. \ref{fig:MR_rho}), too. Since the hadronic part of the EoS is softer than the quark one, the maximal neutron star mass is higher for smaller $\bar\rho$. With decreasing $\bar\rho$ also the radius corresponding to the maximal mass neutron star is increasing. So in this model the parameter set: $g_v = 2.5, \Gamma=0-1, \bar\rho=2\rho_0-3\rho_0$ and the nuclear TNI2u interaction (with K=250 MeV compressibility) fulfills the M-R constraints of neutron star observations.

\begin{figure}
\centerline{\includegraphics[width=0.49\linewidth]{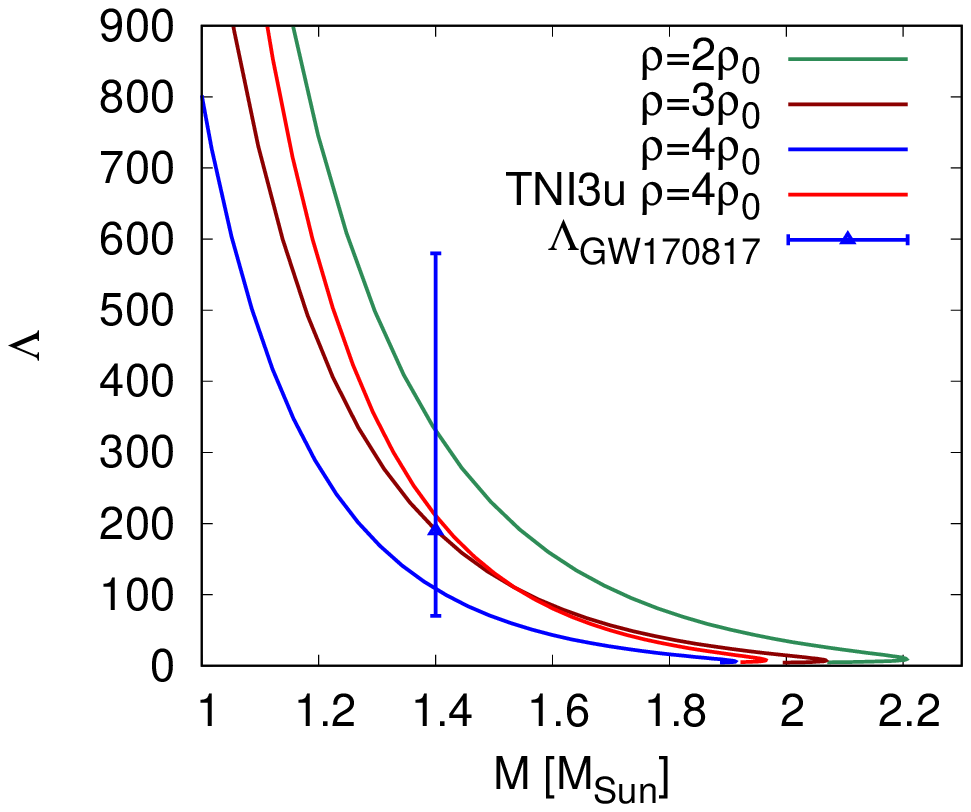}
\includegraphics[width=0.49\linewidth]{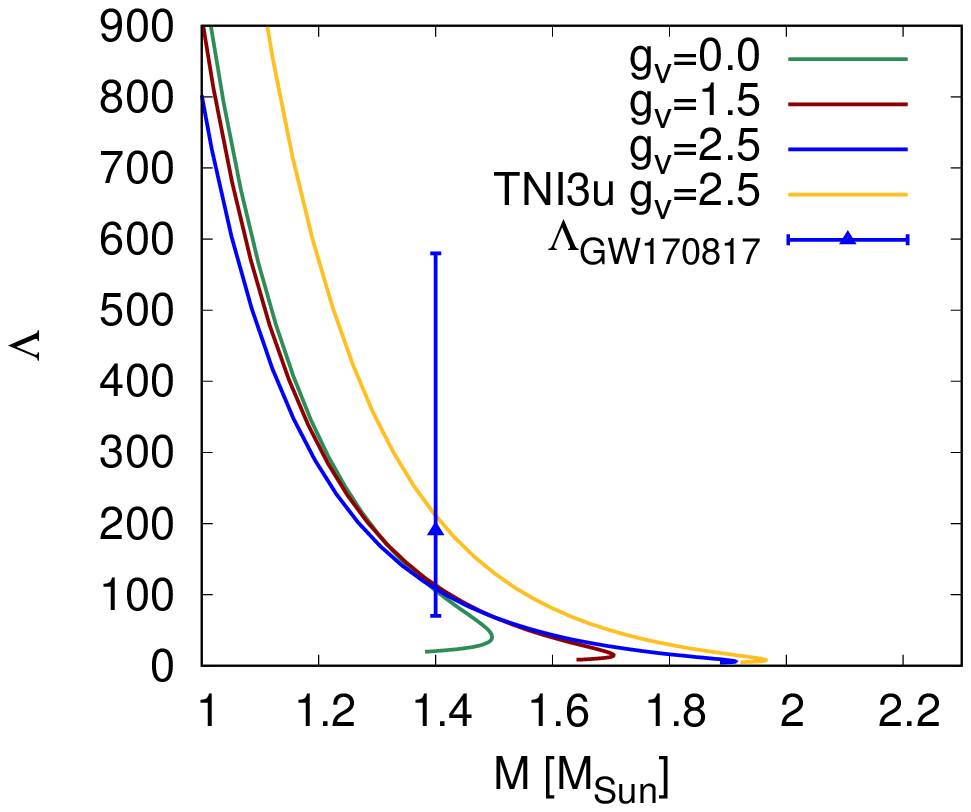}}
\caption{Tidal deformability parameter, $\Lambda$ as a function of the neutron star mass for various crossover densities with $g_v = 2.5, \Gamma=1$ (left), and various coupling strengths between the quarks and vector mesons, $g_v$s with $\Gamma=1, \bar\rho=4\rho_0$ (right). The experimental point is from \cite{GW170817:2018}.}
\label{fig:TidalDef}
\end{figure}

The tidal deformability parameter $\Lambda$ as a function of the neutron star mass is shown for various parameter sets in Fig. \ref{fig:TidalDef}. Since the experimental constraint, $70 < \Lambda_{1.4M_\odot}< 580$ is not very conclusive it does not constrain our parameterizing. However, more precise estimate of $\Lambda$ could rule out models and parameter sets.

\section{Conclusions}  \label{sec:concl}

We employed the zero-temperature EoS to determine the mass-radius relation of compact stars. Including the repulsive interaction in the eLSM model makes the EoS stiff enough to support in some narrow range of the Yukawa coupling compact stars with masses larger than $2$~$M_\odot$ and in the permitted radius window of $11.0$-$12.5$~km at $M=2$~$M_\odot$.

In the future, we would like to go beyond the mean-field approximation, used for the mesons in the eLSM, in a way that takes into account the effect of fermions in the mesonic fluctuations. At lowest order, this can be done by expanding to quadratic order the fermionic determinant obtained after integrating out the quark fields in the partition function and performing the Gaussian integral over the mesonic fields. In order to have a physically more reliable description, we also plan to improve the treatment of the interaction between vector mesons and quarks employed here. Furthermore, to reduce the uncertainty emerging from the artificially performed phase transition, we plan to derive the phase transition parameters from the model itself.

\section{Acknowledgment}
P.~Kov\'acs acknowledges support by the J\'anos B\'olyai Research Scholarship of the Hungarian Academy of Sciences. This work was supperted by an NKFIH-OTKA grant K131982.


\begin{thebibliography}{00}

\bibitem{Demorest:2010}
  Demorest, P.B. et al., Prog. Theor. Phys.Nature 467 (2010) 1081--1083.
\bibitem{GW170817:2018}
  Abbott, B.P. et al., Phys. Rev. Lett. 119 (2017") 161101.
\bibitem{Tlusty:2018rif}
  Tlusty, D., 13th Conference on the Intersections of Particle and Nuclear
  Physics (CIPANP 2018) Palm Springs, California, USA, May 29-June 3 (2018);
      eprint:1810.04767.
\bibitem{Kekelidze:2014dta}
  Kekelidze, V. et al., Proceedings, 2nd International Conference on New
  Frontiers in Physics (ICNFP 2013): Kolymbari, Crete, Greece,
  August 28-September 5, 2013, EPJ Web Conf. 71 (2014) 00127.
\bibitem{Ablyazimov:2017guv}
  Ablyazimov, T. et al., (CBM Coll.), Eur. Phys. J. A53 (2017) 60
\bibitem{lattimer2007}
  Lattimer, J.M. and Prakash, M., Physics Reports 442 (2007) 109--165
\bibitem{Watts:2016uzu}
  Watts, Anna L. et al., Rev. Mod. Phys. 88 (2016) 2
\bibitem{Tews:2018chv}
  Tews, I., Margueron, J. and Reddy, S., Phys. Rev. C98 (2018) 045804.
\bibitem{Tolman}
  Tolman, Richard C., Phys. Rev. 55 (1939) 364--373.
\bibitem{OppVolk}
  Oppenheimer, J. R. and Volkoff, G. M., Phys. Rev. 55 (1939) 374--381.
\bibitem{Hell:2014xva}
  Hell, Thomas and Weise, Wolfram, Phys. Rev. C90 (2014) 045801.
\bibitem{lattimer2014},
  Lattimer, J.M. and Steiner A.W. , The Astrophys. Journal 784 (2014) 123.
\bibitem{kovacs2016}
  Kovacs, P., Szep, Zs. and Wolf, Gy., Phys. Rev. D93 (2016) 114014.
\bibitem{parganlija2013}
  Parganlija, D., Kovacs, P., Wolf, Gy., Giacosa, F. and Rischke, D.H.,
  Phys. Rev. D87 (2013) 014011.
\bibitem{Baym:1971pw}
  Baym, G., Pethick, C. and Sutherland, P., Astrophys. J. 170 (1971) 299--317.
\bibitem{Negele:1973}
  Negele, John W. and Vautherin, Dominique, Nucl. Phys. A207 (1973) 298--320.
\bibitem{TNIu:2006}
  Takatsuka, T., Nishizaki, S., Yamamoto, Y. and Tamagaki, R.,
      Prog. Theor. Phys. 115 (2006) 355--379.
\bibitem{Masuda:2013}
  Masuda, K., Hatsuda, T. and Takatsuka, T., Astrophys. J. 764 (2013) 1.
\bibitem{walecka}
  Walecka, J.D., Annals of Physics 83 (1974) 491--529.
\bibitem{PDG}
  Patrignani, C. et al (PDG), Chin. Phys. C40 (2016) 100001.
\bibitem{james1975}
  F. James and M. Roos, Minuit, Computer Physics Comm. 10 (1975) 343-- 367.
\bibitem{aoki2006}
  Aoki, Y., Fodor, Z., Katz, S.D. and Szabo, K.K., Phys. Lett. B643 (2006) 46--54.
\bibitem{Bazavov:2011nk}
  Bazavov, A. et al., Phys. Rev. D85 (2012) 054503.
\bibitem{hessels2006}
  Hessels, J.W.T. et al., Science (2006) 1901--1904.
\bibitem{almasi}
  Almasi, G., PhD (2017) TU, Darmstadt.
\bibitem{Asakawa:1997}
  Asakawa, M. and Hatsuda, T., Phys. Rev. D55 (1997) 4488.
\end{thebibliography}

\end{document}